\documentclass[a4paper]{spie}  

\usepackage[a4paper,top=2.54cm,bottom=4.94cm,left=1.93cm,right=1.93cm]{geometry} 
\usepackage{amsmath,amsfonts,amssymb}
\usepackage{graphicx}
\usepackage[colorlinks=true, allcolors=blue]{hyperref}
\usepackage{caption}
\usepackage{subcaption}

\title{Low temperature proton irradiation with DEPFETs for Athena's Wide Field Imager}

\author[a]{Valentin Emberger}
\author[a]{Robert Andritschke }
\author[b]{Parviz Azhdarzadeh}
\author[a]{Günter Hauser}
\author[a]{Astrid Mayr}
\author[a]{Johannes Müller-Seidlitz }
\author[b]{Abbas Rezaei}
\author[b]{Wolfgang Treberer-Treberspurg}
\affil[a]{Max-Planck-Institute for Extraterrestrial Physics, Giessenbachstr 1, 85748 Garching, Germany }
\affil[b]{Univ. of Applied Sciences Wiener Neustadt, Johannes-Gutenberg-Straße 3, 2700 Wiener Neustadt, Austria}

\authorinfo{Send correspondence to V. Emberger}

\begin{document} 
\maketitle

\begin{abstract}
The Wide Field Imager (WFI), one of two instruments on ESA's next large X-ray mission
Athena, is designed for imaging spectroscopy of X-rays in the range of 0.2 to 15 keV with a large field of view and high count rate capability. The focal plane consists of back-illuminated DEPFET (Depleted p-channel field effect transistor) sensors that have a high radiation tolerance and provide a near Fano-limited energy resolution. To achieve this, a very low noise readout is required, $\sim$3 electrons ENC at beginning of life is foreseen. This makes the device very susceptible to any radiation induced worsening of the readout noise. The main mechanism of degradation will be the increase of dark current due to displacement damage caused primarily by high energy protons. \\
To study the expected performance degradation, a prototype detector module with fully representative pixel layout and fabrication technology was irradiated with 62.4 MeV protons at the accelerator facility MedAustron in Wiener Neustadt. A total dose equivalent to 3.3 $\cdot$ 10\textsuperscript{9} 10-MeV protons/cm\textsuperscript{2} was applied in two steps. During, in-between and after the irradiations the detector remained at the operating temperature of 213 K and was fully biased and operated. Data was recorded to analyze the signal of all incident particles. \\
We report on the increase of dark current after the irradiation and present the current related damage rate at 213 K. The effect of low temperature annealing at 213\,, 236\,K, 253\,K, 273\,K, and 289\,K is presented. 
\end{abstract}


\section{INTRODUCTION}
\label{sec:intro}  

The focal plane of the WFI\cite{arnewfi} instrument for Athena\cite{paul13, Ayre17} will consist of DEPFET sensors. The current design and recent results of electrical and performance characterization can be found in Refs.~\citenum{mibo19, mibo22, nom20}.  DEPFET sensors  are active pixel  sensors. One advantage over state of the art CCD-based X-ray sensors is that they are not depending on the charge transfer efficiency (CTE). In space the CTE suffers from the radiation induced creation of traps, if these traps are in that region on the surface of the device where the charge transfer occurs. The DEPFET sensors can therefore be considered more radiation hard and can possibly maintain their energy resolution throughout the mission. To achieve that goal the operating temperature has to be low enough to keep the radiation induced increase of dark current at an acceptable level. As all defects that are created in the sensitive volume of the sensor can contribute to the increase of dark current this effect is particularly relevant for thick X-ray sensors. The DEPFET sensors for Athena's WFI have a 450 µm thick fully depleted bulk in order to achieve a sufficient quantum efficiency in the targeted energy range from 0.2 to 15 keV. Therefore the increase of leakage current is likely the dominant effect of performance degradation due to radiation. 

There are two basic mechanisms to mitigate this. The first is shielding the device from charged particles. Shielding against high energy protons is however limited due to mass and space constraints. Furthermore, the use of a massive proton shield increases the number of secondary particles. These can deposit an amount of energy in the sensor that lies within the range of interest and therefore these events contribute to the instrumental background. The focal plane of the WFI instrument is currently foreseen to have a proton shield equivalent to 40 mm Aluminum in all directions other than the optical path towards the mirror. Simulations that predict the Total Ionizing Dose (TID), the Total Non Ionizing Dose (TNID) and the instrumental background\cite{tanja21} have been performed for this configuration. It is expected that during the goal life time of 10 years and in an orbit around L2 the sensors of the WFI will receive a TNID of 1.3 $\cdot$ 10\textsuperscript{9} 10-MeV protons/cm\textsuperscript{2}.

The second way to mitigate the noise increase is to cool the detector. At lower temperature the current generation rate of defects is decreased. This is limited by the available radiator area for passive cooling of the various sub-systems of WFI. The aim of this study is to determine the radiation induced leakage current at the predicted TNID and its temperature dependence.

The relation of TNID and leakage current for silicon sensors has been studied by many groups, the most intensive studies were done for sensors in particle detectors i.e. the RD48 (ROSE) collaboration. But the temperature of the device during irradiation is important because of annealing effects. Furthermore, surface related effects that depend on the structures on the front side can play a role and lead to non uniform dark current distributions and hot pixels. Thus, to make precise predictions for new devices it is best to perform dedicated measurements that take the operating conditions into account. 

\section{THEORY}
\label{sec:theory}  

The WFI instrument is supposed to achieve an end-of-life energy resolution of 170 eV at 7 keV and 80 eV at 1 keV. A very low readout noise of $\sim$3 electrons Equivalent Noise Charge (e\textsuperscript{-}  ENC) across the complete readout chain is needed to fulfill that requirement. There are three main contributors to the total read noise, detector, Detector Electronics (DE) and Ground Segment (GS):
\begin{equation} \label{eq:tot_noise}
\sigma_{total}^2 = \sigma_{detector}^2 + \sigma_{DE}^2 +\sigma_{GS}^2 
\end{equation}
These can be divided into numerous sub components, estimations for each have been derived from measurements or analysis.  A detailed budget is presented in Ref.~\citenum{astrid24}.  Here we focus only on the detector noise that is composed of three components:
\begin{equation} \label{eq:det_noise}
\sigma_{detector}^2 = \sigma_{signal}^2 + \sigma_{depfet}^2 +\sigma_{asic}^2
\end{equation}
The signal noise is the shot noise due to the thermal generation of charge carriers that occurs in the silicon bulk of the sensor. It is a crucial component because it is expected to experience the strongest change during the mission. It is equal to the square root of the number of generated electrons per readout cycle.
\begin{equation} \label{eq:sig_noise}
\sigma_{signal}  = \omega\sqrt{g\cdot t} \, ,
\end{equation}
with frame time t, generation rate g and mean electron hole pair creation energy $\omega$. Here $\omega$ = 3.71 eV is used.

In orbit the signal noise increases due to lattice distortions that are introduced through particle radiation, i.e. the Total Non Ionizing Dose (TNID). The TNID that results from a particle depends on the particle type and energy. In the scope of the NIEL scaling theory they can be compared to each other under the assumption that the damage to the lattice is proportional to the Non Ionizing Energy Loss (NIEL) of the particle. It is then possible to express a certain TNID as a displacement damage equivalent fluence (DDEF) that is defined as follows:
\begin{equation} \label{eq:ddef}
\Phi_{eq} = \kappa \: \Phi_{p,E} \, , 
\end{equation}
where $\Phi_{eq}$ is the equivalent flux, $\kappa$ the so called hardness factor and $\Phi_{p,E}$ the flux of particles of type p and energy E. The equivalent flux is usually defined for neutrons with an energy of 1 MeV. In this work the equivalent flux of 10-MeV protons is also used.
The hardness factor $\kappa$ is given by:
\begin{equation} \label{eq:hardness}
\kappa = \frac{NIEL(E)_{p}}{NIEL(E_{eq})_{p_{eq}}} \, ,
\end{equation}
where $NIEL(E)_{p}$ is the non ionizing energy loss of a particle with type p and energy E, $E_{eq}$ and $p_{eq}$ are the energy and particle type for which the equivalent flux is defined.

The relation of a TNID dose and the resulting increase of dark current depends on the specific device and its annealing history. It is expressed as the current related damage rate $\alpha$:
\begin{equation}\label{eq:damage_rate}
\Delta I = \alpha \: \Phi_{eq(n, 1MeV)} \: V \, ,
\end{equation}
where $\Delta I$ is the increase in dark current and V is the irradiated volume.

As the dark current results from thermal excitation of charge carriers it shows a very strong temperature dependence. It is usually given by 
\begin{equation} 
\label{eq:Tdep}
\frac{I_{dark}(T_1)}{I_{dark}(T_2)} = \left( \frac{T_1}{T_2} \right)^2\:e^{\frac{E_g}{2k_B} \left( \frac{1}{T_1}-\frac{1}{T_2} \right) } \, ,
\end{equation}
 where $k_B$ is the Boltzmann constant. 
At room temperature the parameter E\textsubscript{g} is usually assumed to be identical to the band gap energy of 1.12 eV, but there are cases where lower values are found to produce a better agreement with the measurement data. A possible explanation for this is given in Ref.~\citenum{darkCCD}. The authors argue that at low temperatures changing contributions from depletion dark current and diffusion dark current can lead to a temperature dependence of E\textsubscript{g}.  

\section{PARTICLE ACCELERATOR}
\label{sec:accel}  

The irradiation took place at the synchrotron facility MedAustron in Wiener Neustadt, Austria. It is an irradiation facility that is mainly used for medical purposes. The accelerator can deliver proton fluxes of up to  $10^9$~p\textsuperscript{+}/s  with energies from 62.4\,MeV to 800\,MeV.\\
There is one beamline that is reserved for scientific research.  At this beam line three special low flux settings~\cite{pur21} are available, that allow to set the particle rates in a range from a few kHz to several MHz. 
For these low flux settings the built-in dosimetry system located in the beam exit window is however not available. 
The synchrotron delivers spills of protons with a duration of 5 s followed by a break of 4 s. Within the spills the flux can have variations with unspecified amplitude and frequency, therefore flux values are only given as mean flux.\\
The particle energy as given by the accelerator is defined  at the location of the so called isocenter that is located in the irradiation room, 85 cm away from the beam exit window. Inside the exit window there is material equivalent to 2.4 mm of water, thus there will be an energy dispersion of the protons at that spot. This is however not specified. Only the peak energy is given.   

\section{Experiment}
\label{sec:setup}  

\begin{figure} [ht]
   \begin{center}
   \begin{tabular}{c} 
   \includegraphics[height=5cm]{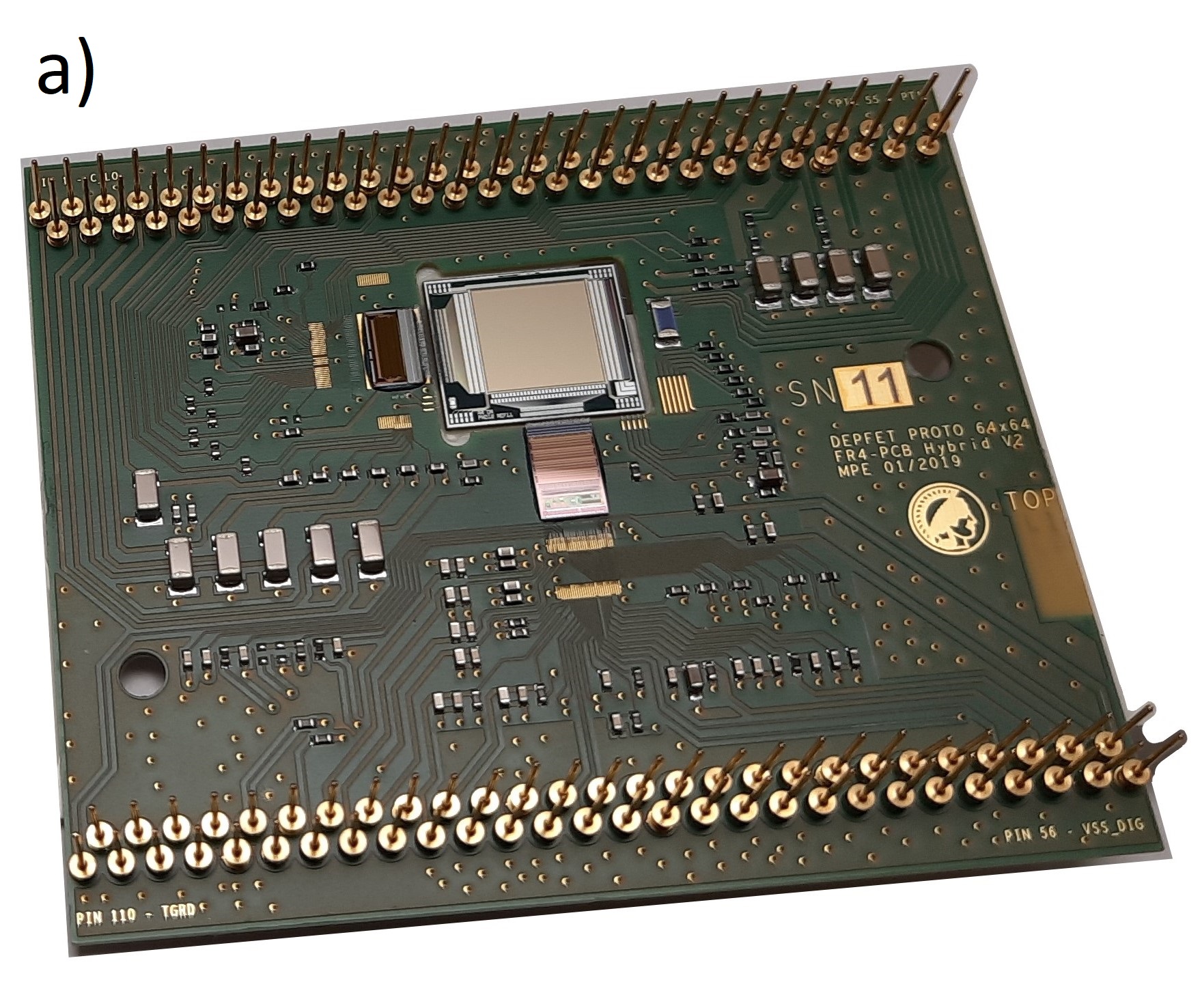}
   \includegraphics[height=5cm]{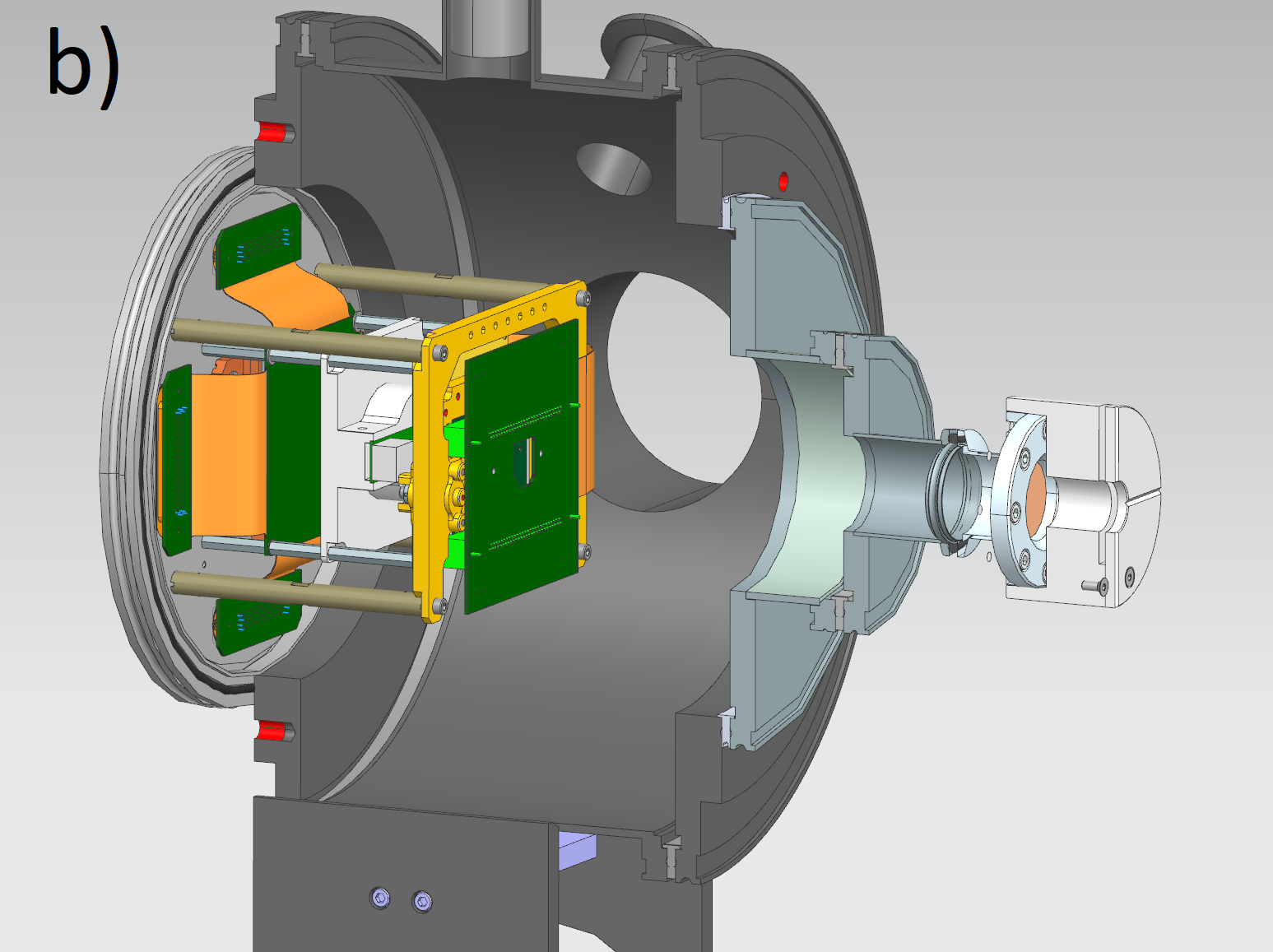}
   \includegraphics[height=5cm]{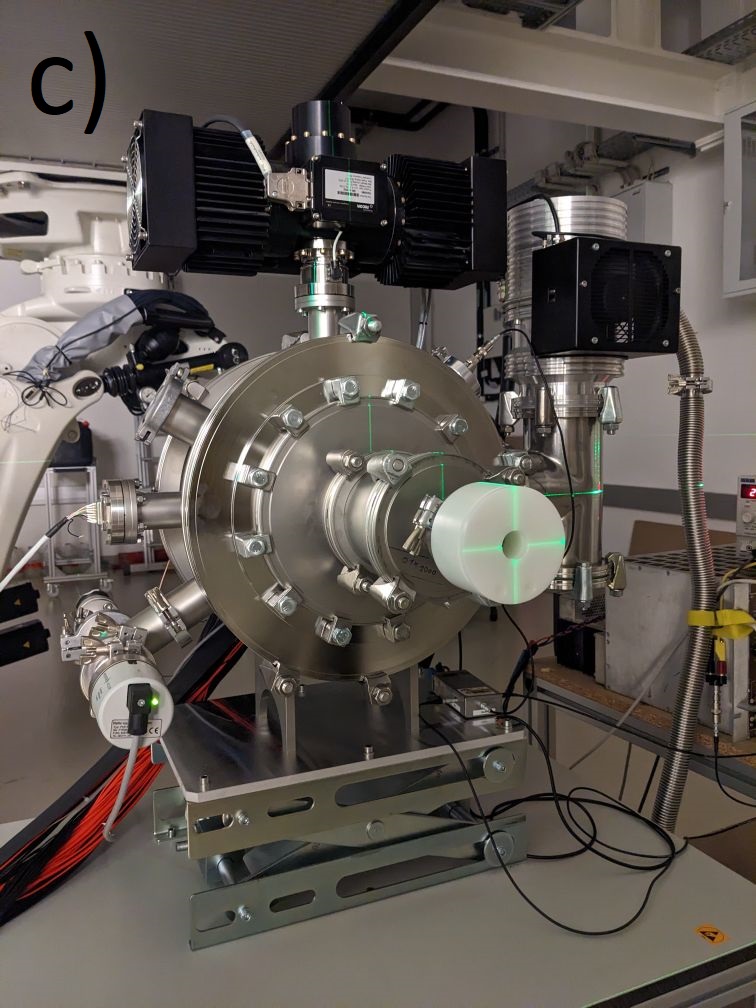}
	\end{tabular}
	\end{center}
   \caption[Setup] 
   { \label{fig:chamber} 
a) Detector module with 64x64 pixel DEPFET sensor, VERITAS readout ASIC and Switcher A steering ASIC. b) Vacuum chamber with measurement setup for irradiating the front side of the sensor and the ASICs. c) Positioning was done with the laser positioning system available in the irradiation room.}
   \end{figure}

The detector module that was used for the irradiation test has a 64$\times$64 pixel DEPFET sensor glued to a ceramic carrier. This carrier is attached to a PCB that holds a readout ASIC (VERITAS 2.2~\cite{sven18, anna24}), a steering ASIC (Switcher A\cite{fischer03}) and some passive components (Figure~\ref{fig:chamber}a). The pixels of the DEPFET sensor are identical to those of the Athena flight sensors in terms of layout and process technology. \\ 
The vacuum chamber is shown in Figure~\ref{fig:chamber}b. It is equipped with a stirling cooler that brings the entire detector module to the operating temperature that is specified for the sensor. For power supply, steering and readout there are feed troughs on the rear side of the chamber. These are connected via a flexible lead to an adapter board that holds the detector module. This adapter board has an opening that is large enough to irradiate the front side and the ASICs without any additional material in the way. The cover of the chamber is equipped with a dedicated proton entrance window that needs to be very thin but light tight. It is composed of only 50\,µm of polyimide with a nickel coating and thus has a minimal proton energy degradation (\textless 70 keV @ 70 MeV). \\
In order to avoid activation of any high-Z materials of the setup during irradiation a collimator and a beam stop were designed. The collimator was located in front of the entrance window and the beam stop was located in the chamber behind the detector module, both consisting of Polytetrafluoroethylene (Teflon) with a thickness of 4 cm (Figure~\ref{fig:chamber}b). 

The positioning was done by aligning the front edge of the setup and a cross on the collimator to the laser positioning system in the irradiation room (Figure \ref{fig:chamber}c).\\
The total allocated beam time was two times 8 hours with a break of 24 hours in between. The goal was to keep the flux low enough to enable dosimetry with the DEPFET sensor itself and at the same time high enough to reach at least the expected dose during the goal life time of 10 years.\\
To avoid any unwanted annealing the detector has to stay at the same temperature during the two irradiation periods. Therefore the setup was equipped with an uninterruptible power supply. After the irradiations it was moved to the decay room located next to the irradiation room. This room is not accessible at all times and has no possibility for data connection. Therefore the system was setup to autonomously monitor the detector noise in between the two irradiation slots. Based on the observed annealing during this period it was decided that after receiving the full dose the detector shall be kept at the irradiation temperature for at least a few weeks  before the final measurement is done.

\section{MEASUREMENT RESULTS}
\label{sec:result}  

\subsection{Dosimetry}
\label{sec:dosimetry}
In this experiment the dosimetry relies completely on the detection of the radiation with the device under test. With an average energy deposition of 876 keV each proton produces a very clear signal in the DEPFET sensor but the signal is beyond the targeted energy range that reaches only until 15 keV. By changing the gain of the preamplifier  the range can however be extended to about 445 keV without significant changes in the operating parameters. Solely the DEPFET current was decreased from 100 µA per channel to 25 µA per channel in order to also reduce the gain of the DEPFET itself. With this setting the number of saturated pixels per event is reduced to one or two for most proton hits while quadruple events with almost equal charge sharing of the four pixels have no saturated pixels and therefore allow even a measurement of the energy deposit of the incoming protons. The detector was read out with a frame time of 160 µs. \\
With these settings the protons can be counted using the standard event filtering and recombination algorithms as they are described in Ref.~\citenum{robert08} and Ref.~\citenum{lauf14}. This is however limited to situations where the flux stays low enough so that pile up of events can be neglected. This is true for the average flux of protons during the complete irradiation (about  4.1$\cdot$10\textsuperscript{5} 62.4-MeV~protons/cm\textsuperscript{2}$\cdot$s). But if the flux variations are taken into account not all frames can fulfill this condition as will be shown in the following.

\subsubsection{Mean signal per event}
\begin{figure} [ht]
   \begin{center}
   \begin{tabular}{c} 
   \includegraphics[width=\textwidth]{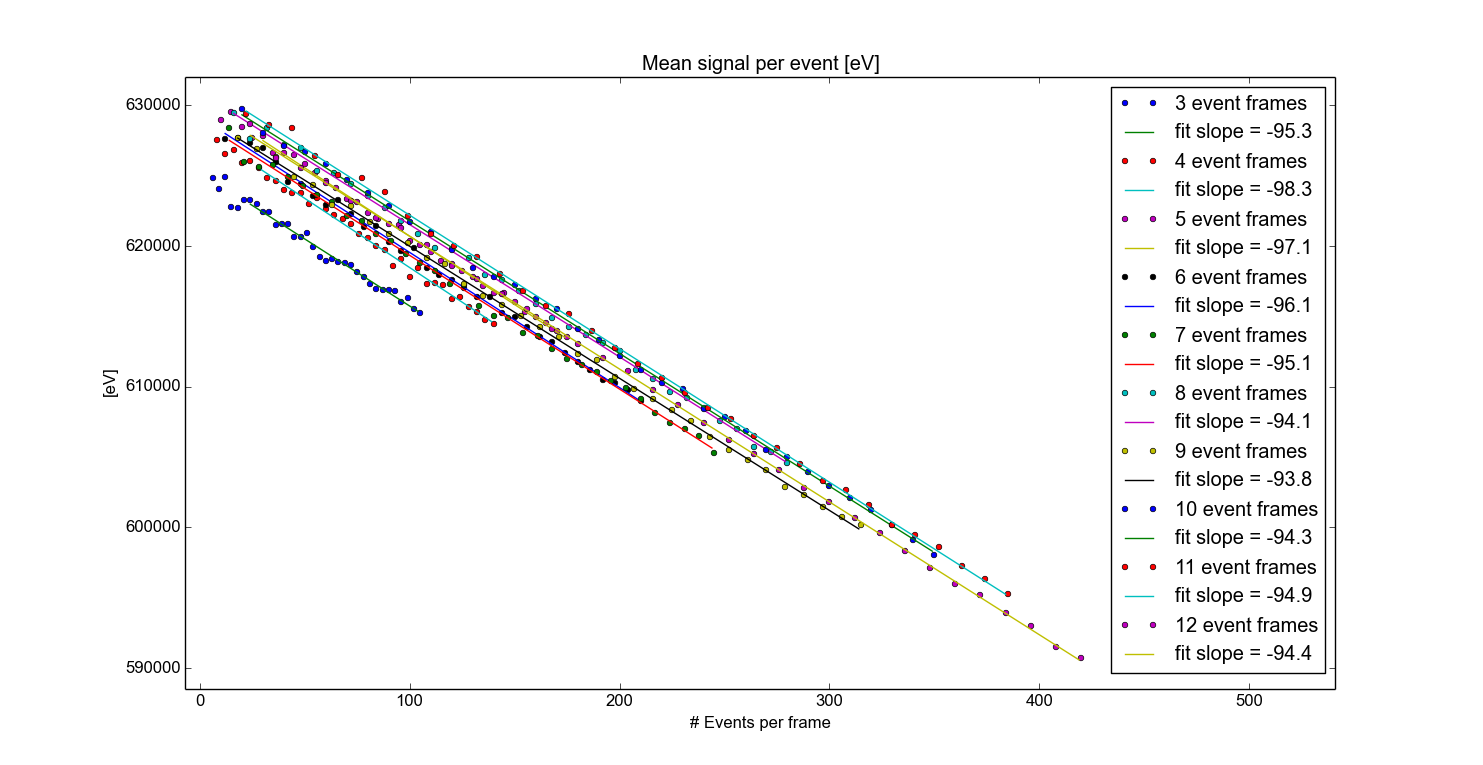}
	\end{tabular}
	\end{center}
   \caption[] 
   { \label{fig:slope_event} 
The mean signal per event depends on the number of events per frame due to the increasing overlap of event patterns and the consequent loss of signal in saturated pixels.}
   \end{figure}

\begin{figure} [ht]
   \begin{center}
   \begin{tabular}{c} 
   \includegraphics[width=\textwidth]{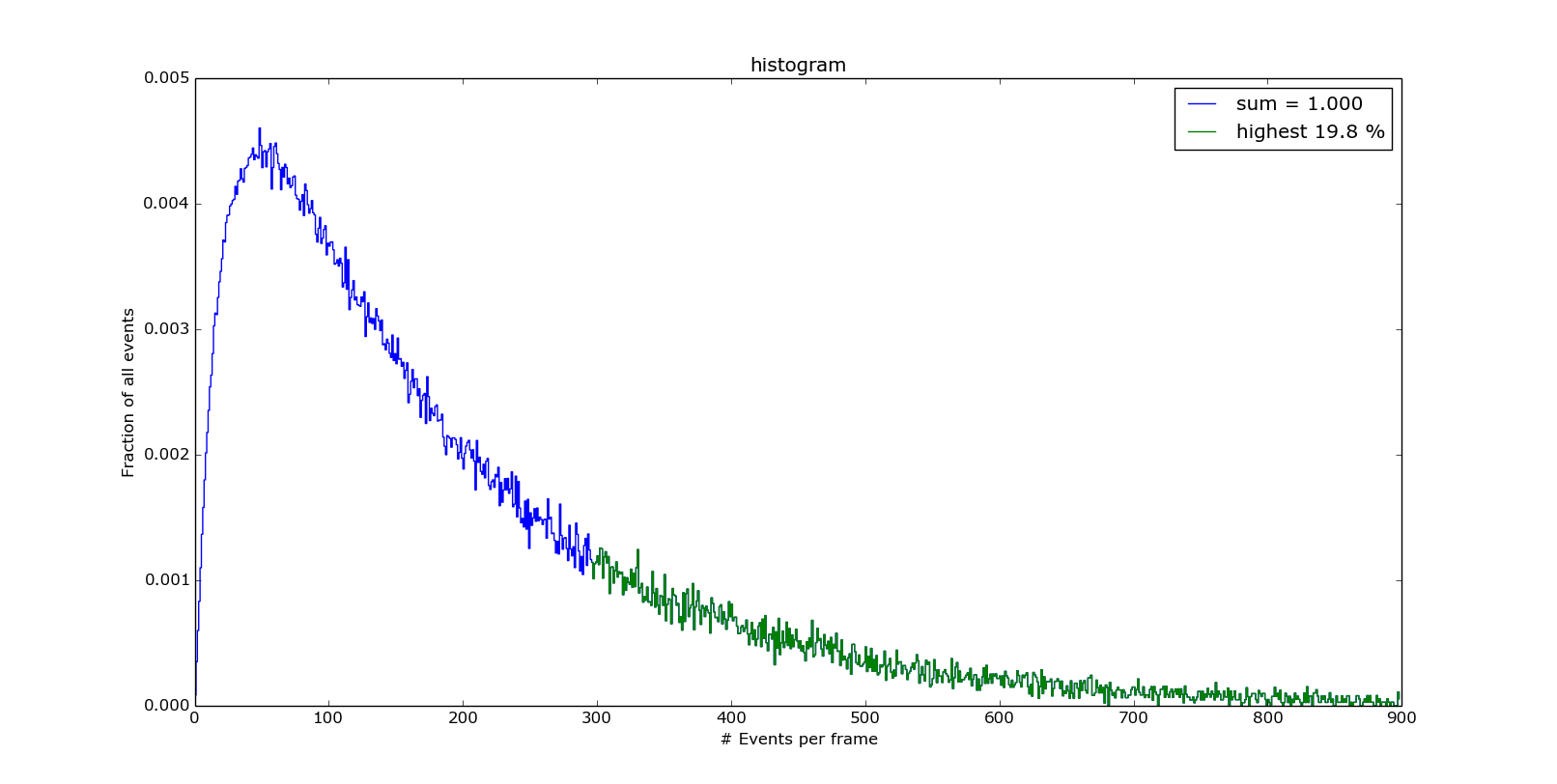}
	\end{tabular}
	\end{center}
   \caption[Mean signal per event] 
   { \label{fig:Intensity_hist} 
The distribution of the number of events per frame illustrates the extent of the flux variations.}
   \end{figure}

A different approach to count the number of protons is by finding the mean signal per proton event in dependence of the number of protons per frame. To calculate this value a number of frames with only a few events (3 to 12) was selected from the data set. From these the mean signal per event can be calculated with very low contamination of those cases where pileup occurs. The mean signal per frame should however change at high fluxes because of the increasing overlap of event patterns and the consequent loss of signal in saturated pixels. In order to check how the mean signal per event scales with the number of protons per frame the above mentioned frames with just a few events can be stacked in arbitrary combinations while the maximal signal is capped at the saturation value (Figure~\ref{fig:slope_event}). 
The mean signal per event is found to be (628.8\,\textpm\,4.7)\,keV, at sufficiently high fluxes decreasing by (95.3\,$\pm$\,1.4)\,eV$\cdot$N, where N is the number of protons per frame.
In a similar manner also the dependence of the mean signal per event on the total signal per frame can be derived. This relation then allows to calculate the number of protons per frame without any event detection and recombination algorithms and is therefore unaffected by pileup. This number may not be accurate for each individual case, but should be adequate to study the flux distribution in a statistical manner. 
The distribution (Figure \ref{fig:Intensity_hist}) shows that there is a significant number of events that originate from frames that show such a high intensity that reliable proton counting is not possible anymore.

\subsubsection{Dose and dose rate}
\label{sec:dose}

\begin{figure} [ht]
   \begin{center}
   \begin{tabular}{c} 
   \includegraphics[width=\textwidth]{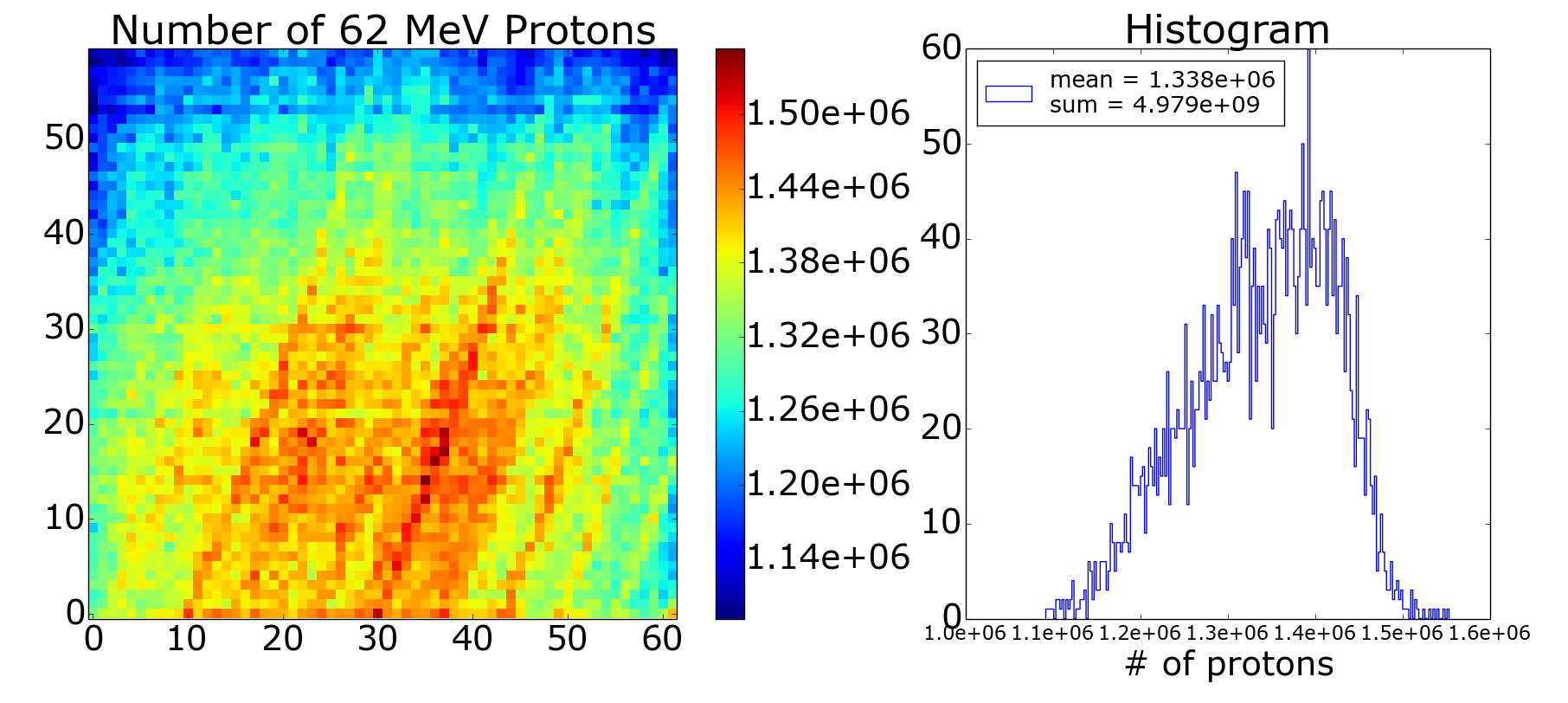}
	\end{tabular}
	\end{center}
   \caption[Number of protons] 
   { \label{fig:proton_map} 
Map of the number of protons per pixel. The FWHM of the beam profile is clearly larger than the sensor.}
   \end{figure}

Due to the unexpected high flux in part of the frames it was chosen to calculate the total dose by using the above mentioned relations for the mean signal per event. The total number of protons in a pixel results from the sum over the signal of this pixel in all frames divided by the mean signal per event. In frames with very high intensity the signal is corrected for the loss of signal in saturated pixels. This correction however increases the resulting fluence  by only 3\%. For the conservative lower limit estimation of the fluence it was therefore omitted. The result is shown in Figure~\ref{fig:proton_map}. The origin of the diagonal stripes is not understood yet and needs further investigation. As the relative intensity variation across these stripes is only a few percent it is not considered that they change the final conclusion significantly.    

The proton fluence is converted to a DDEF of 10-MeV protons as given by Equation~\ref{eq:ddef}  with the hardness factor as defined in Equation~\ref{eq:hardness}.    

\begin{figure} [ht]
   \begin{center}
   \begin{tabular}{c} 
   \includegraphics[width=\textwidth]{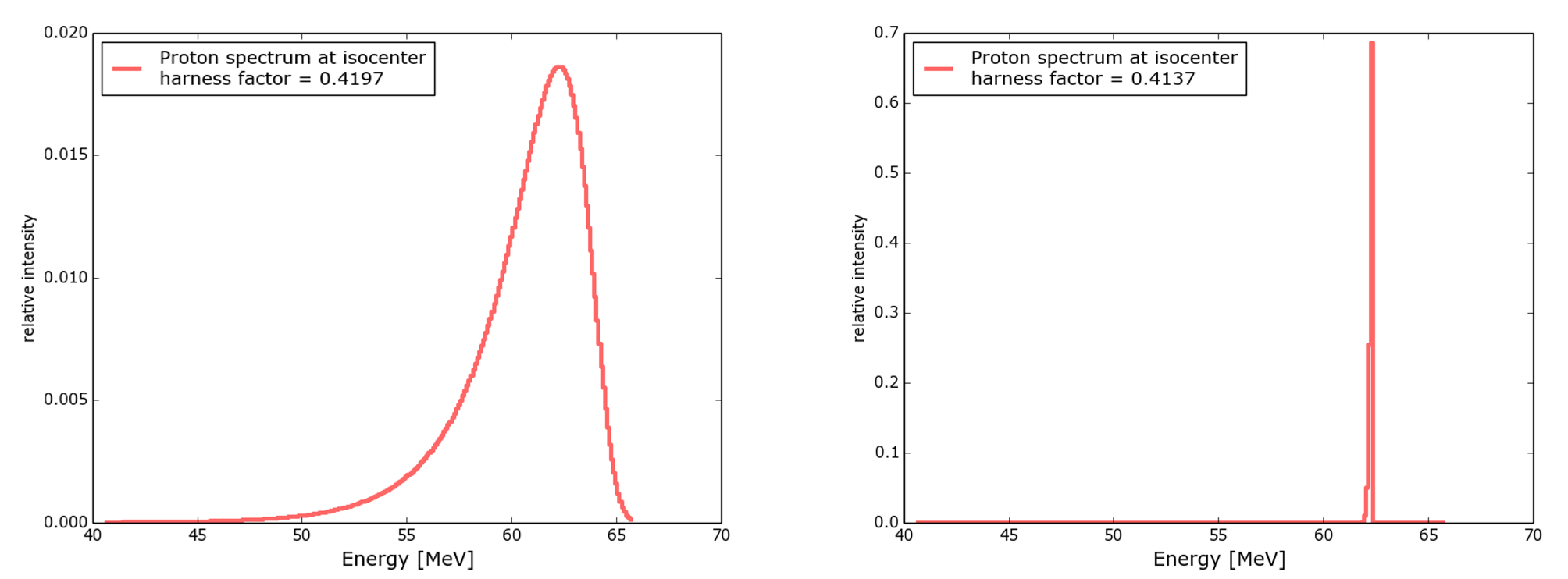}
	\end{tabular}
	\end{center}
   \caption[Number of protons] 
   {\label{fig:prot_spec} 
Hardness factors calculated from simulated proton spectra with a peak energy of 62.4 MeV. The energy loss is a moyal distribution with a peak at 3.41 MeV and a width of 1.3 (left) or 0.03 MeV (right).  }
   \end{figure}

The passage of the protons through the beam exit window (equivalent to 2.4 mm of water) and 85 cm of air leads to an average energy loss of 3.41 MeV until they reach the isocenter with a peak energy of 62.4\,MeV. As a result of the energy straggeling the protons have an energy spectrum of certain shape and width. The effective hardness factor $\kappa_\text{eff}$ in principle depends on the shape of the spectrum. But as the width will be less than a few MeV it is not necessary to calculate it in detail. In a first order approximation the NIEL in a range of a few MeV around 62.4\,MeV is sufficiently linear and the shape of the spectrum sufficiently symmetric that the effective hardness factor will not deviate significantly from that of mono energetic protons with the peak energy of 62.4\,MeV. To illustrate this we computed two proton spectra with a peak energy of 62.4\,MeV where the energy loss is a moyal distribution with 3.41\,MeV peak and a width of 0.03\,MeV or 1.3\,MeV (see Figure~\ref{fig:prot_spec}). These spectra were weighted with tabulated stopping power values from Ref.~\citenum{niel}. The resulting hardness factors differ by less than 1\%. For the dose calculation the hardness factor $\kappa_\text{eff} = 0.4167\,\pm\,0.0030$ was used.

A mean dose equivalent to 3.3$\cdot$10\textsuperscript{9}~10-MeV~protons/cm\textsuperscript{2} was applied. With the total irradiation time of 19,570\,s (not counting the 4 s break between the 5 s long spills)  we find a mean dose rate of 1.7$\cdot$10\textsuperscript{5}~10-MeV~protons/cm\textsuperscript{2}$\cdot$s. According to the distribution of events per frame (see Figure~\ref{fig:Intensity_hist}), the peak dose rate should be on the order of 3$\cdot$10\textsuperscript{6}~10-MeV~protons/cm\textsuperscript{2}$\cdot$s. 

\subsection{Dark Current}
\label{sec:dark current}
\begin{figure} [ht]
   \begin{center}
   \begin{tabular}{c} 
   \includegraphics[width=0.5\textwidth]{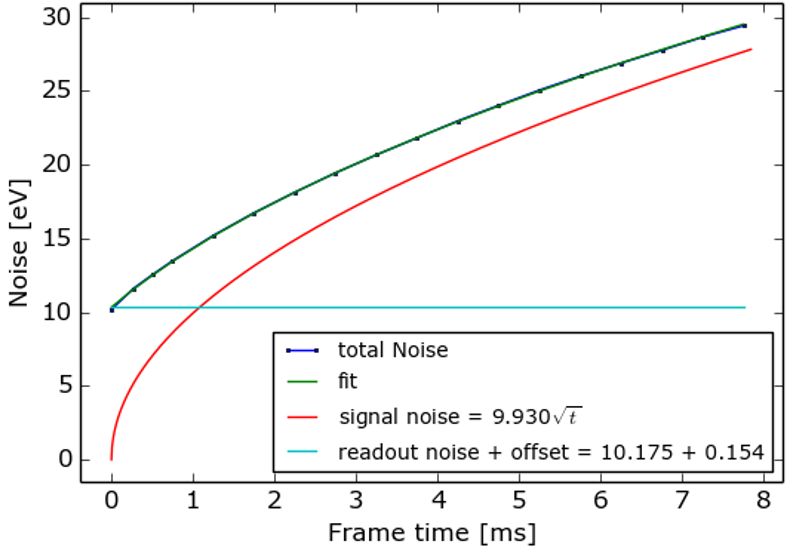}
  \includegraphics[width=0.5\textwidth]{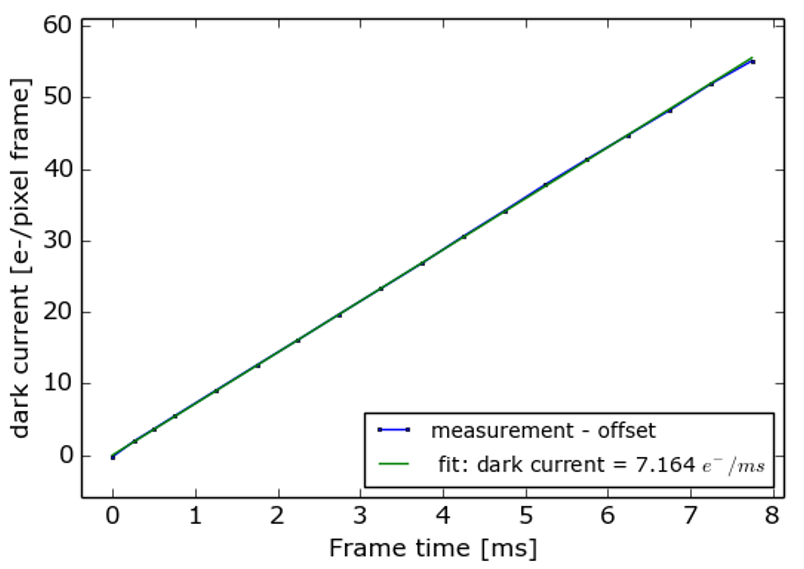}
	\end{tabular}
	\end{center}
   \caption[Number of protons] 
   { \label{fig:dark_curr_fit} 
Dependence of the mean noise on the exposure time fitted with a constant and a frame time dependent signal noise component. The constant is the sum of the first measurement point and an offset that is derived from the fit. The resulting dark current per pixel and ms is shown in the right graph. The measurement was done at $\approx$\,214 K.}
   \end{figure}

\begin{figure} [ht]
   \begin{center}
   \begin{tabular}{c} 
   \includegraphics[width=\textwidth]{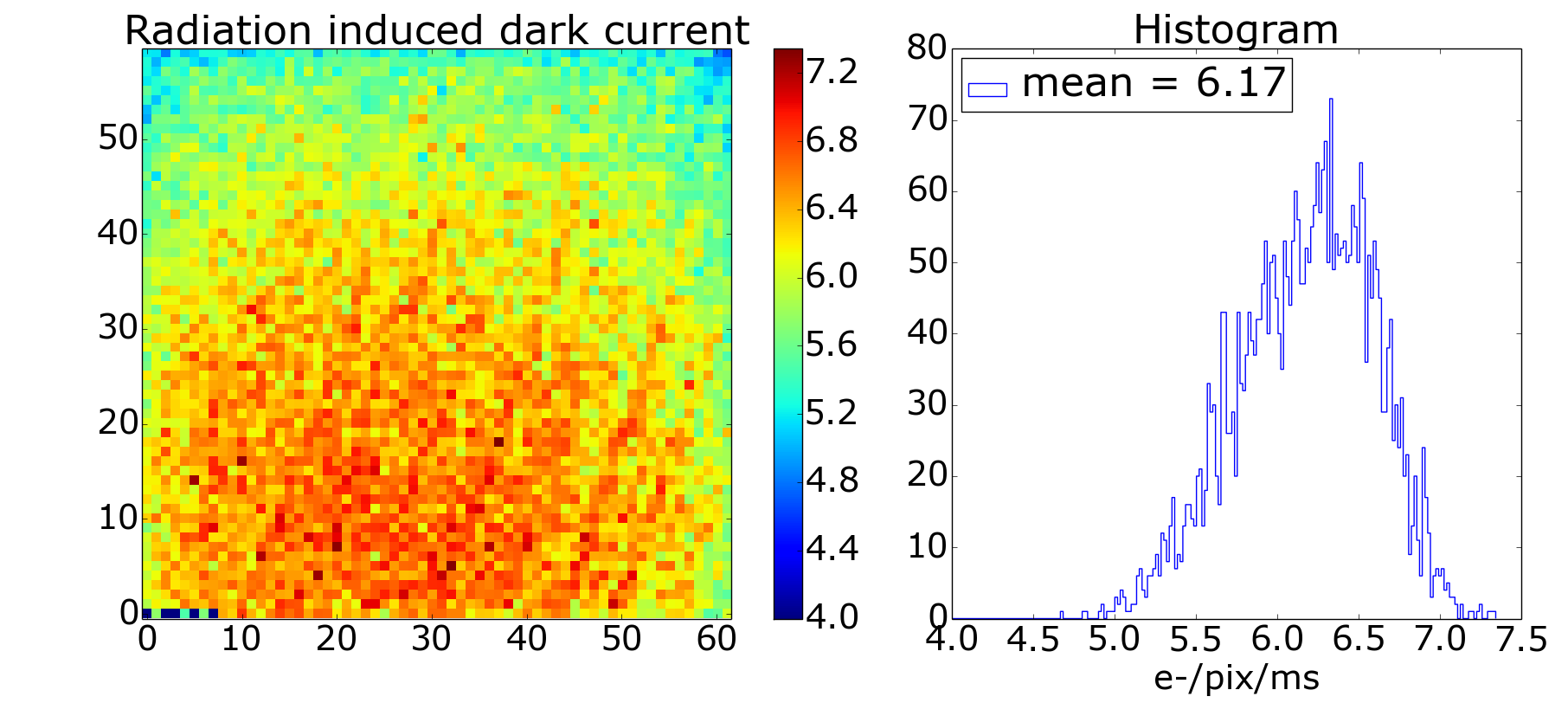}
	\end{tabular}
	\end{center}
   \caption[Number of protons] 
   { \label{fig:dark_curr_map} 
Radiation induced dark current at 213 K. Derived through interpolation from measurements at four different temperatures and subtraction of the pre-irradiation dark current. }
   \end{figure}

The dark current of each pixel was measured before, in-between and after the irradiation. To measure it, the exposure time for each frame was varied in the range from 2\,µs to 7.5\,ms and 1000 dark frames were saved at each setting. The measurement with the shortest frame time was realized by using a special control sequence where all charges are cleared from the internal gate directly before the readout. Noise maps are calculated from these dark frames. The dependence of the noise on the exposure time is fitted according to Equation~\ref{eq:det_noise} and \ref{eq:sig_noise} to get the dark current. Figure~\ref{fig:dark_curr_fit} shows the fit on the mean noise of all pixels. In Figure~\ref{fig:dark_curr_map} the result of the pixelwise analysis is shown. The measurement was taken 30 days after the end of the irradiation during which the detector stayed at the same temperature as during the irradiation (213\,K).
 
\subsection{Temperature Dependence and Annealing}
\label{sec:temp_dep}

\begin{figure} [ht]
   \begin{center}
   \begin{tabular}{cc}
   \includegraphics[width=0.5\textwidth]{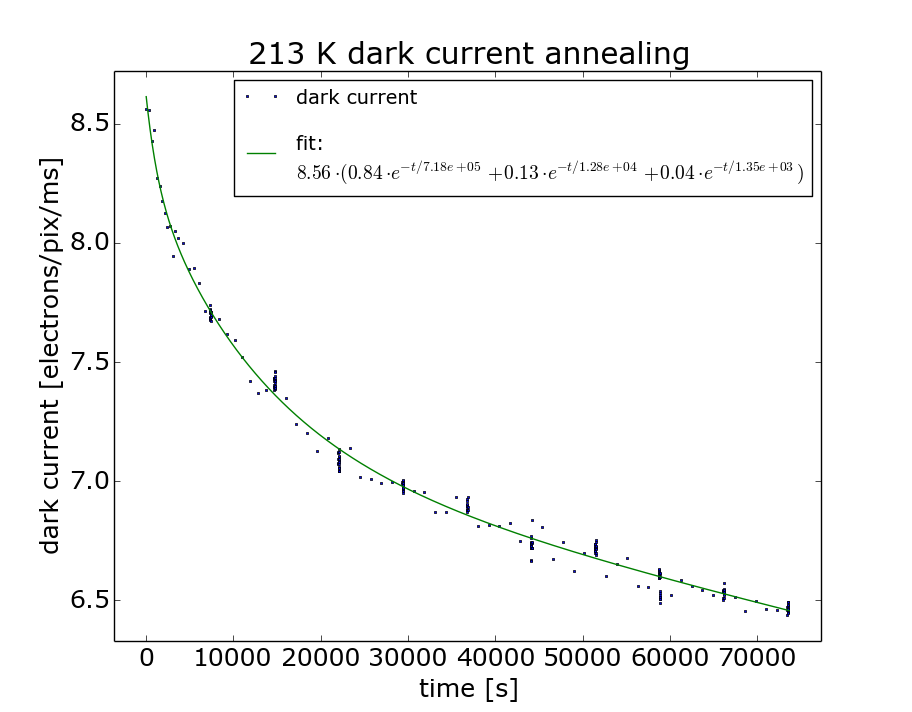}
   \includegraphics[width=0.5\textwidth]{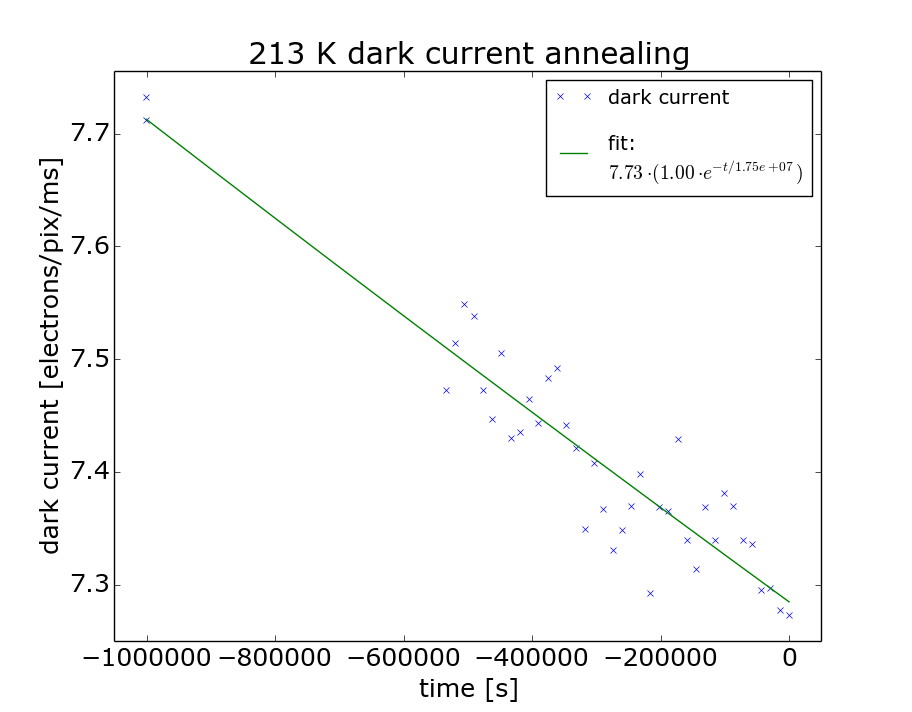}
	\end{tabular}
	\end{center}
   \caption[] 
   { \label{fig:int_anneal} 
Left: Twenty hours of annealing at 213 K after the first irradiation slot. Right: Ten days of annealing at 213 K starting about 20 days after the irradiation.}
   \end{figure}

In room temperature irradiation experiments the annealing of defects and the resulting decrease of dark current is significant. For this reason the dose rate and the annealing history need to be considered when analyzing such an experiment which makes the results difficult to compare. Therefore it is common to apply a defined period of annealing at a higher temperature before the measurement. Due to the strong temperature dependence of the annealing rate all other room temperature annealing that happened during irradiation or before the high temperature annealing is then insignificant. A 80/60 annealing (80 minutes at 60\,°C) is often used as reference. 
Most of the annealing as it is observed at room temperature is connected to defect clusters who are not mobile at a temperature of 213 K because their activation energy is too high. Therefore their annealing is frozen in our experiment. The primary defects and simple complexes like divacancies and some vacancy-impurity pairs are however still mobile at much lower temperatures. In room temperature experiments they anneal rapidly or form more stable complexes. This is often referred to as short term annealing and occurs within minutes after the irradiation. Therefore in most experiments it is not even observed. At 213 K this "short term" annealing happens on much longer timescales. This is illustrated on the left side of Figure \ref{fig:int_anneal} where the time evolution of the dark current after the first irradiation slot is shown. A this point about 1/3 of the final dose was applied. The best fit on the annealing curve could be achieved using three time constants. The largest component has a half life of 12 days. Thus it is clear that a measurement that predicts the radiation induced dark current of the sensor in space cannot be done directly after the irradiation. Consequently the detector was kept at the operating temperature for 30 days. No data is available from the first 20 days of this cold annealing period, but some noise measurements were done during the last 10 days. From these measurements (Figure~\ref{fig:int_anneal}, right) it can be seen that the dark current is still measurably decreasing. Even though this data is not enough to reliably determine the annealing rate it can be seen that the time constant should be at least in the order of 10\textsuperscript{7} seconds. Thus, the measurement 30 days after the irradiation can serve as a conservative estimation of the dark current after several years in orbit.        

\begin{figure} [ht]
   \begin{center}
   \begin{tabular}{cc} 
   \includegraphics[height=6cm]{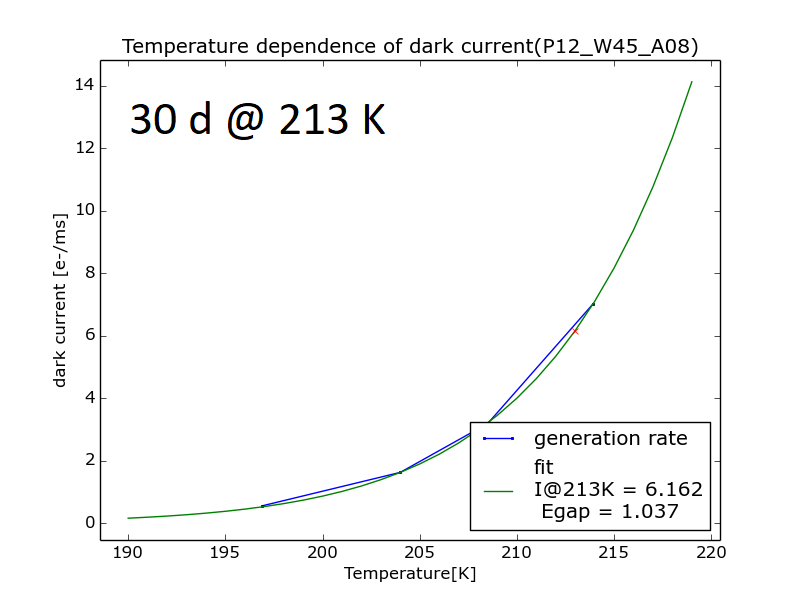}& \includegraphics[height=6cm]{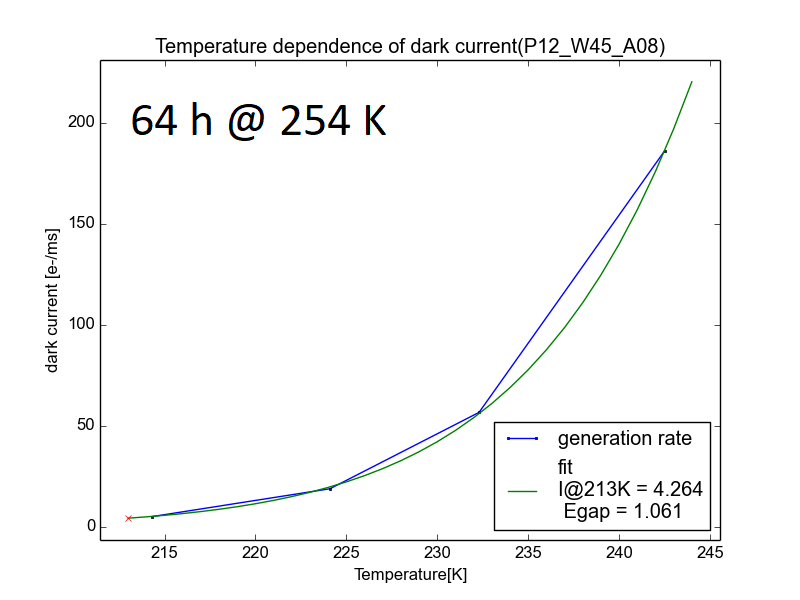}\\
   \includegraphics[height=6cm]{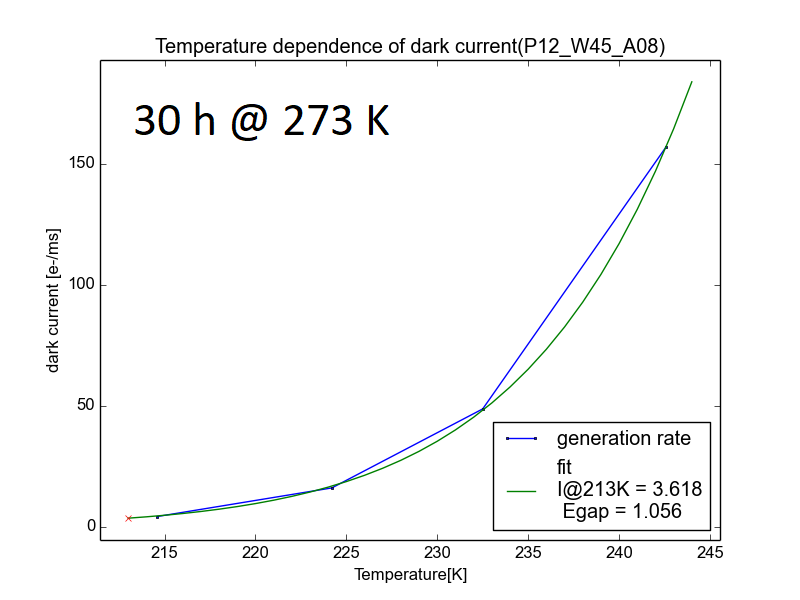}& \includegraphics[height=6cm]{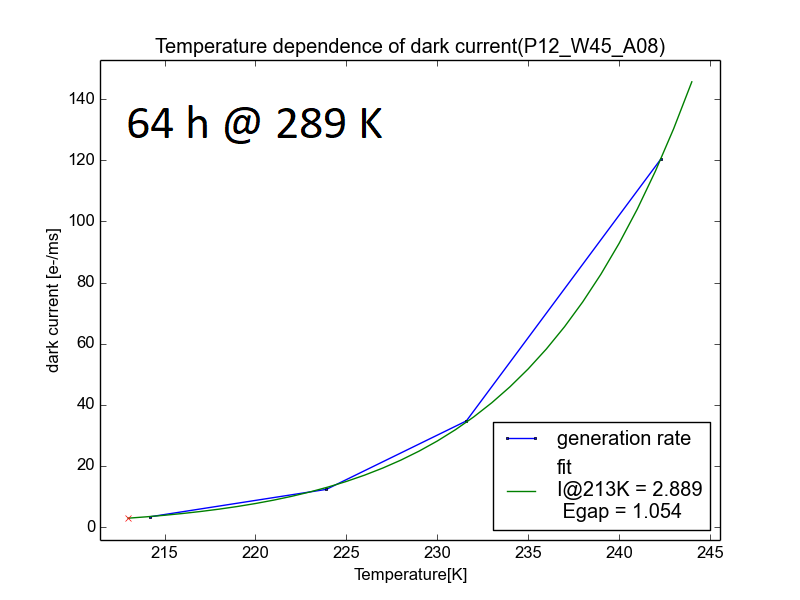}
	\end{tabular}
	\end{center}
   \caption[] 
   { \label{fig:temp_dep} 
Temperature dependence of dark current after the long annealing at 213\,K and after three other annealings at higher temperatures.}
   \end{figure}

After the first measurement of the dark current at the irradiation temperature the measurement was repeated at three lower temperatures. The resulting dark currents were fitted according to Equation~\ref{eq:Tdep} to get the dark current at exactly 213\,K.  For the temperature dependence of dark current E\textsubscript{g} = 1.12\,eV is most often used. For the temperature range from 197\,K to 214\,K the measured temperature dependence was better reproduced with a slightly lower value. Therefore the free parameters in the fit are the dark current at 213\,K and E\textsubscript{g}. The result of the fit is shown in Figure~\ref{fig:temp_dep}.
After these first measurements potential annealing scenarios were studied by annealing the device at four different slightly to moderately elevated temperatures for a certain amount of time. After each annealing step the procedure of deriving the dark current at 213\,K was repeated. All annealing steps and the resulting values of the mean dark current are summarized in Table~\ref{tab:anneal}. 

It is remarkable that even at these comparably low annealing temperatures the dark current reduced significantly. An annealing procedure in orbit at a temperature below 300\,K appears to be possible and beneficial. It is observed that the readout noise of the DEPFETs is increased during the annealing steps which hints at the formation of interface traps at the Si/SiO\textsubscript{2} interface on the structured side of the device. This is in accordance with the results from a previous TID test\cite{vale22}. It cannot be evaluated on the basis of this experiment in how far these interface traps contribute also to the remaining dark current after the annealing. A further TID test is planned to clarify this.  

\begin{table}[ht]
\caption{Dark current and DEPFET noise after five annealing steps with varying temperature and duration. The dark current at 213 K was interpolated from four measurements at the temperatures T1 to T4. Note that the dark current decreases while the DEPFET noise increases. } 
\label{tab:anneal}  
\begin{center}       
\begin{tabular}{|l|l|l|l|l|l|l|l|} 
\hline
\rule[-1ex]{0pt}{3.5ex}  Annealing & Annealing & Meas.  & Meas. & Meas.  & Meas.  & I\textsubscript{dark}@213 K & DEPFET \\
\rule[-1ex]{0pt}{3.5ex} Temp. [K] & Time &  T1 [K]  & T2 [K] & T3 [K] & T4 [K] & [e\textsuperscript{-}/ms$\cdot$pix] & noise [eV] \\
\hline
\rule[-1ex]{0pt}{3.5ex} Pre irrad. & n.a. & 214 &  &  & & 0.16 & 7.88 \\
\hline
\rule[-1ex]{0pt}{3.5ex} 213 &	30 d & 214 & 197 & 204 & 208 & 6.22 & 8.73 \\
\hline
\rule[-1ex]{0pt}{3.5ex} 236 &	16 h & 236 & 213 & 218 & 223 & 6.09 & 8.90 \\
\hline
\rule[-1ex]{0pt}{3.5ex} 254 &	64 h & 214 & 224 & 232 & 243 & 4.26 & 9.34 \\
\hline
\rule[-1ex]{0pt}{3.5ex} 271 &	30 h & 214 & 224 & 232 & 243 & 3.62 & 9.91 \\
\hline
\rule[-1ex]{0pt}{3.5ex} 289 &	64 h & 214 & 224 & 232 & 243 & 2.89 & 10.43 \\
\hline
\end{tabular}
\end{center}
\end{table} 

\section{DISCUSSION}
\label{sec:discuss}  
\subsection{Current Related Damage Rate}
\label{sec:damage_rate}  

By definition the current related damage rate is calculated with the dark current at 20\,°C and by using the equivalent fluence of 1 MeV neutrons. Thus,  the currents measured at 213 K should be scaled according to Equation~\ref{eq:Tdep}. For the scaling E\textsubscript{g} is assumed to be 1.12 eV in order to make the results comparable. With the measured equivalent fluence  $\Phi_{eq(p^{+}, 10 MeV)}=3.3\cdot$10\textsuperscript{9}\,10-MeV~protons/cm\textsuperscript{2} , the ratio of the NIEL of 1 MeV neutrons  and 10-MeV~protons~(3.871), the pixel dimensions of 450\,µm\,$\times$\,130\,µm\,$\times$\,130\,µm (sensitive volume) and  the measured increase of dark current of 6.22\,e\textsuperscript{-}/pix$\cdot$ms at 213\,K we calculate the current related damage rate of silicon
\begin{center}
$\alpha =  \frac{\Delta I}{\Phi_{eq(n, 1 \mathrm{MeV})}} \: V = 8.2 \: (+0.6 - 0.4) \cdot 10^{-17}$\,A/cm  \end{center}
for irradiation at 213 K and subsequent annealing of 30 days at 213 K. The error includes the error estimates on the dose and the dark current, but no error on the E\textsubscript{g} parameter for the temperature scaling.

G. Segneri et al.~\cite{mixs_rate} have done a similar experiment in preparation of the ESA mission Bepi Colombo. They irradiated diodes with 10 MeV protons at 223 K. They reported a significantly higher damage rate of $\alpha = 1.11 \pm 0.02 \cdot 10^{-16}$\,A/cm.  The discrepancy can be explained by the long annealing time at 213 K before the measurement. With an assumed time constant of 1$\cdot 10^7$ s (see Section~\ref{sec:temp_dep}) it can be interpolated that shortly after the irradiation the result would have been $1.06\cdot 10^{-16}$\,A/cm which is already in accordance with the above mentioned result without considering any further processes with shorter time constants.

\subsection{Consequences For the Mission }
\label{sec:mission}  
On the base of the results we present a preliminary prediction for the end-of-life (EOL) performance of the WFI detectors and its temperature dependence. For the sake of simplicity, we restrict the calculations here to the mean values for all sensor pixels. It has to be noted that for a final conclusion the pixel individual doses, dark currents and other noise components with their respective distributions have to be accounted for if requirements are to be met for all individual pixels. \\
For the following results the performance budget and the requirements from the WFI systems requirements review are used. These could be subject to change in the scope of the successful reformulation of the Athena mission.

\begin{figure} [ht]
   \begin{center}
   \begin{tabular}{cc}
   \includegraphics[width=0.46\textwidth]{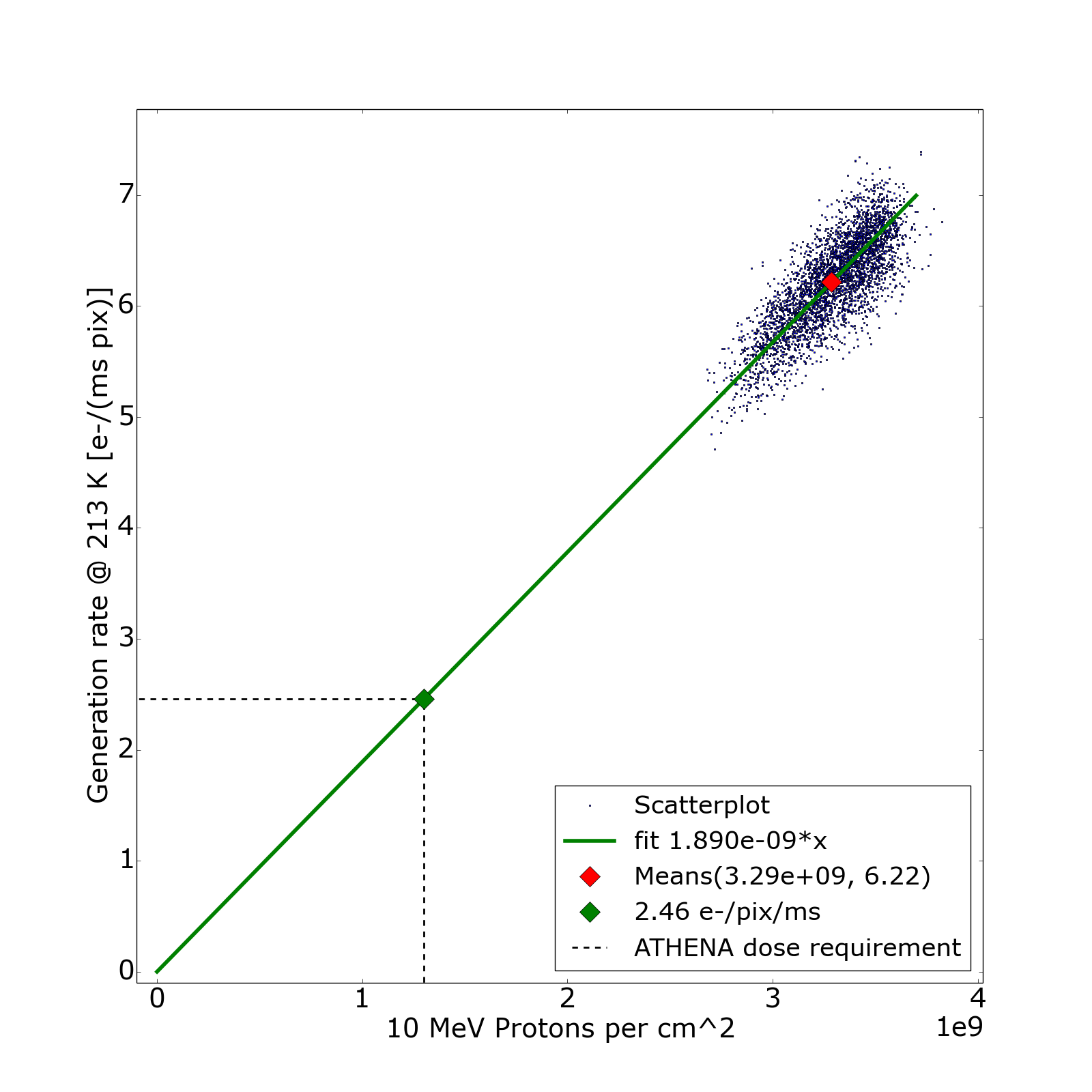}
   \includegraphics[width=0.54\textwidth]{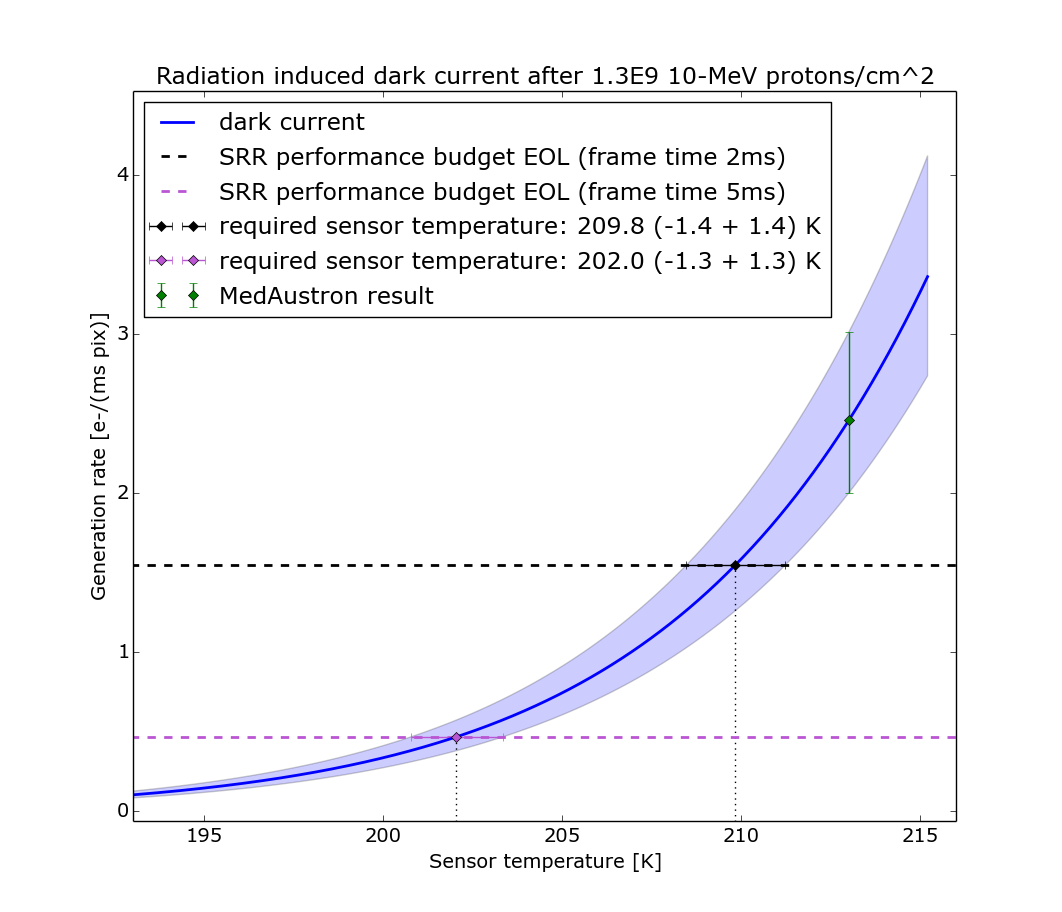}

	\end{tabular}
	\end{center}
   \caption[] 
   { \label{fig:dose_curr} 
Left: Scatterplot of the dose and dark current values of all pixels. The green dot represents the expected dark current at the expected dose after 10 years in space and thus the main result of this experiment. Right: The green dot from the left figure is plotted with error bar that results from the uncertainty of the dose and current measurements. The dashed lines show the dark current budgets for a frame time of 2 and 5 ms, the dotted lines the necessary sensor temperature for that value.}
   \end{figure}

\begin{figure} [ht]
   \begin{center}
   \begin{tabular}{cc}
   \includegraphics[width=0.5\textwidth]{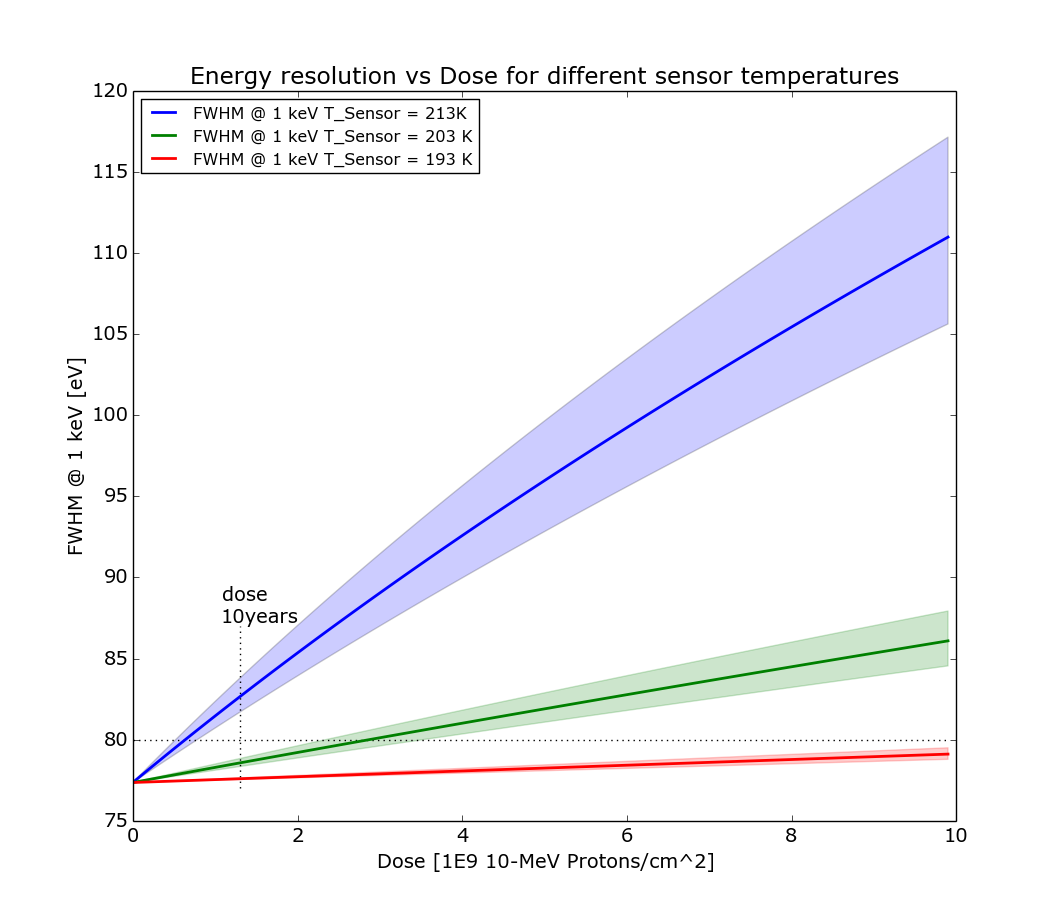}
   \includegraphics[width=0.5\textwidth]{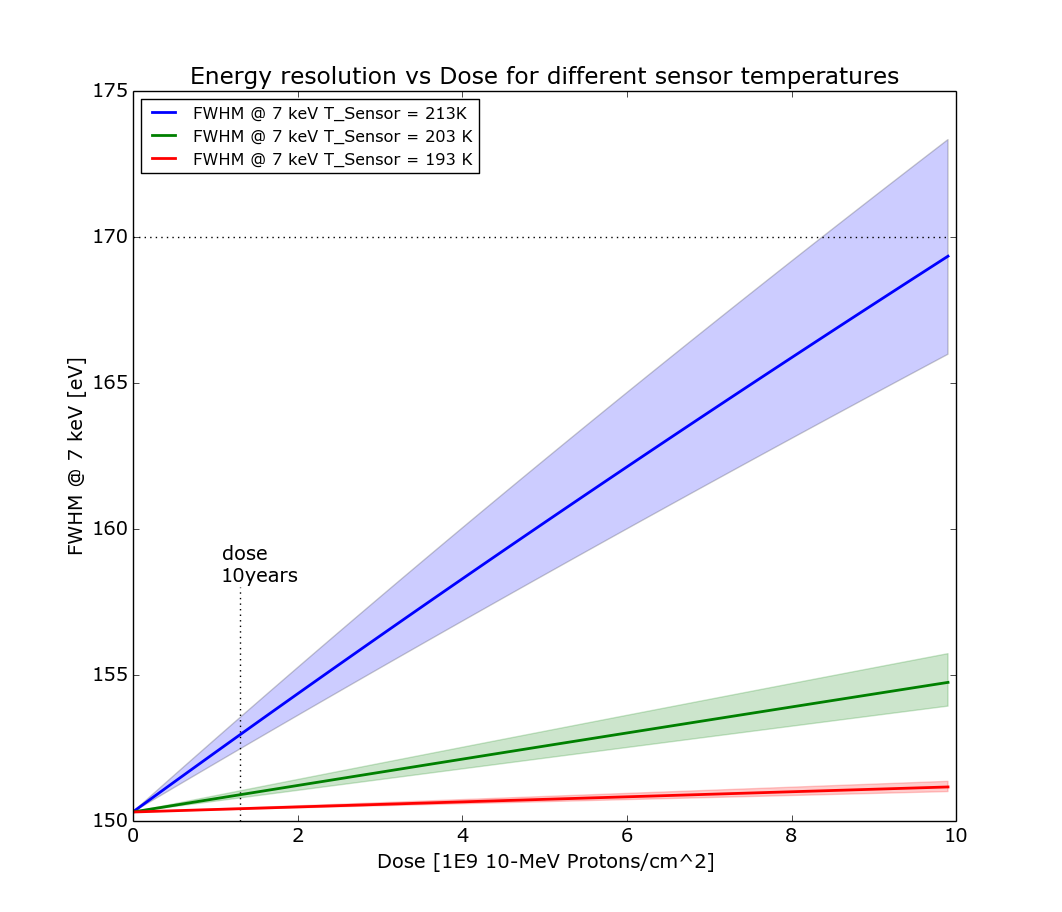}
	\end{tabular}
	\end{center}
   \caption[] 
   { \label{fig:fwhm_dose} 
A calculation illustrates the evolution of the energy resolution with the accumulated TNID and the associated increase of leakage current for three different sensor temperatures. On the left it is shown for a line energy of 1 keV, on the right for 7 keV.}
   \end{figure}

Figure~\ref{fig:dose_curr} shows a scatterplot of the dose and dark current values of all pixels except of those at the border of the device. A linear fit on the points through zero shows the dependence of the radiation induced dark current on the dose, the extrapolated dark current value for the expected dose after the goal life time of ten years is shown in green. The predicted dark current and its temperature dependence can be used to estimate the necessary operating temperature at EOL to stay within the budget for the dark current (Figure~\ref{fig:dose_curr}, right). For a frame rate of 500 Hz the operating temperature needs to be below (209.8~\textpm~1.4)\,K.
Assuming all other noise components at their budgeted values and constant during the mission we can calculate the energy resolution as a function of the dose. In Figure \ref{fig:fwhm_dose} this is shown for 3 exemplary temperatures and line energies of 1\,keV and 7\,keV. Note that these are simplified calculations that serve for illustration. Dedicated simulations for performance prediction can be found in Ref.~\citenum{joschi24}.

\section{SUMMARY}
\label{sec:summary}  
A detector module with a DEPFET sensor from the pre-flight production for Athena's WFI was irradiated with 64.2\,MeV protons at a temperature of 213 K. The TNID was 3.3$\cdot$10\textsuperscript{9}\,10-MeV-protons/cm\textsuperscript{2}. After the irradiation the detector remained at the irradiation temperature for thirty days. The subsequent measurements showed a remaining dark current of  6.2\,$e^-$/ms$\cdot$pixel. Based on this measurement  it is predicted that at after 10 years in orbit (and a dose of 1.3$\cdot$10\textsuperscript{9}\,10-MeV-protons/cm\textsuperscript{2}) the sensor will have a radiation induced dark current of 2.5\,$e^-$/ms$\cdot$pixel at 213\,K. This implies that the operating temperature needs to be below (209.8\,\textpm\,1.4)\,K  in order to keep the signal noise within the budget that was allocated during the system requirements review.

We observed significant annealing at 213\,K during the first weeks after the irradiation but there is not enough data to reliably determine the annealing rate. Further annealing steps with temperatures in the range from 236\,K to 289\,K were done and indicate that an annealing procedure in orbit at temperatures below 300\,K should be possible and beneficial.

 \section*{ACKNOWLEDGMENTS} 
The authors thank Tomas Schreiner from MedAustron for access to and support at the accelerator facility. Development and production of the DEPFET sensors for the Athena WFI is performed in a collaboration between MPE and the MPG Semiconductor Laboratory (HLL). The work was funded by by the Max-Planck-Society and the German space agency DLR (FKZ: 50 QR 1901 and FKZ: 50 QR 2301).
 
\bibliography{Low_T_proton} 
\bibliographystyle{spiebib} 

\end{document}